\documentclass[preprint2]{aastex}

\newcommand{\gapprox}{$\stackrel {>}{_{\sim}}$}   %greater/less approx.

\slugcomment{To appear in Ap.J.}

\shorttitle{IR view of EXors: the case of V1118 Ori}
\shortauthors{Lorenzetti et al.}

\begin{document}

\title{An infrared view of the EXor variables:\\ on the
case of V1118 Ori
\thanks{Based on observations mainly collected at the AZT-24 telescope
(Campo Imperatore, Italy)}}

\author{D.Lorenzetti\altaffilmark{2},
    T.Giannini\altaffilmark{2},
    V.M.Larionov\altaffilmark{3,4},
    E.Kopatskaya\altaffilmark{3},
    A.A.Arkharov\altaffilmark{4},
        M.De Luca\altaffilmark{2,5}
          and
        A.Di Paola\altaffilmark{2}
          }
%\offprints{Dario Lorenzetti, email:dloren@mporzio.astro.it}

\altaffiltext{2}{INAF - Osservatorio Astronomico di Roma, via Frascati 33,I-00040
 Monte Porzio, Italy, dloren, giannini, deluca, dipaola@oa-roma.inaf.it}
\altaffiltext{3}{Astronomical Institute of St.Petersburg University, Russia,
vlar@nm.ru, kopac@EK13929.spb.edu}
\altaffiltext{4}{Central Astronomical Observatory of Pulkovo, Pulkovskoe shosse 65,
196140 St.Petersburg, Russia,arkharov@mail.ru}
\altaffiltext{5}{Universit\`a degli Studi di Roma "Tor Vergata" - Dipartimento di
Fisica, via della Ricerca Scientifica 1, I-00133 Roma, Italy}
%
%\email{dloren, giannini, deluca, dipaola@mporzio.astro.it;
%vlar@nm.ru; kopac@EK13929.spb.edu; arkharov@mail.ru} }
%\date{Received;Accepted}

%\abstract {}{}{}{}
% 5 {} token are mandatory

\begin{abstract}
We investigate the relationship between the infrared observed
properties of the EXor variables and the mechanisms active during
their evolutionary stage typical of the pre-Main-Sequence
phase.

To this aim we have constructed a catalog containing all the
IR (1-100 $\mu$m) photometric and spectroscopic observations
appeared during the last 30 years in the literature.
Moreover, new results of our monitoring program based on near- and
mid-IR photometry, near-IR spectroscopy and polarimetry of one
object (V1118 Ori) typical of the EXor class are presented,
complementing those given in a previous paper and related to a
different activity period.

Our catalog indicates how the database accumulated so far, stemming
from a fortuitous monitoring of the stellar activity, is inadequate
for any statistical study of the EXor events. Nevertheless, all the
observational evidences can be interpreted into a coherent scheme.
The sources that present the largest brightness variations tend to
become bluer while brightening. The scenario of the disk accretion
hypothesis based on the viscous friction between particles agrees
with the observational evidences. The new results on V1118 Ori
confirm such general view. The striking novelty is represented by a
near-IR spectrum of V1118 Ori taken one year later the last monitored outburst: any emission line previously detected is
now totally disappeared at our sensitivity. For the same
source, mid-IR photometry is provided here for the first time and
allows to construct a meaningful spectral energy distribution. The
first polarimetric data indicate the source is intrinsically
polarized and its spotted and magnetized surface becomes
recognizable during the less active phases.
\end{abstract}


%\authorrunning{Lorenzetti D. et al.}
%\titlerunning{IR view of EXOrs: the case of V1118 Ori}

%\maketitle

%
% _______________________________________________________________

\section{Introduction}

Herbig (1989) noticed and listed eight pre-main sequence stars
characterized by repetitive outbursts in optical light, referring
them as EXors, after the example first recognized, EX Lup. Their
outbursts have an amplitude up to 5 magnitudes, lasting one year or
less, with a recurrence time of about 5-10 years. As for the FUor
class, outbursts are explained in terms of disk accretion events
(Hartmann, Kenyon \& Hartigan 1993) which occur during an
evolutionary phase between the main accretion phase and the
beginning of the main sequence evolution. The EXor stage appears to
follow the FUor one, more than being a less evident
manifestation of the same phase (Lorenzetti et al. 2006 - Paper I);
moreover EXors are less luminous than FUors and present
different emission line spectra to those of FUors (which are
dominated by absorption features). However, many observational evidences remain still unexplained: (i) the cause(s) of the short
timescale (hours, day) variability, maybe linked to different time scales in the accretion processes; (ii) the (cor)-relations
between optical-IR and X-ray behaviour, which can elucidate on how
accretion regulates the coronal emission; (iii) whether the observed
shape variations of the spectral energy distribution (SED) are
related with different levels of activity; (iv) the presence and the
influence of close companions.

The above considerations mainly stem from observations in the near- and mid-infrared (1-10 $\mu$m), where the emission of the
accreting disks is expected to peak. The IR studies are 
crucial during both inactive and active phases: in the former
stages they allow us to investigate how the properties of the
circumstellar matter prior the outburst will influence,
through the accretion, the outburst itself; while, in the latter stages the same observations sample how the circumstellar
material is altered by intermittent mass loss.

Unfortunately, both the scanty number of recognized EXors and the
lack of an adequate database of photometric-spectroscopic
multi-frequency observations are the consequence of a widespread
concept, according to which long term monitoring programs are
considered not rewarding; indeed, these projects appear as
inadequate for large telescope equipped with fore-front
instrumentation and remain a prerogative of small telescopes, where,
however, a relevant fraction of astronomers do not invest their
resources. To foster these monitoring programs on EXors, after
having defined the EXors sample in Sect.2, we provide in Sect.3 a
record of all the IR (1-100 $\mu$m) photometric and spectroscopic
observations reported in the last 30 years in the literature;
in Sect.4 the observations of our monitoring of the EXor V1118 Ori
are presented, complementing those already given in Paper I. Sect.5
is dedicated to discuss the results of the presented observations,
while the concluding remarks are given in Sect.6.

\section{The sample}

%%%%   TABLE - OBJECT COORDINATES  %%%%%%%%%%%%%%%%%%%%%%%%%%%%%%%%%%%%%%%%%%%%%%%%
\begin{deluxetable}{ccccc}
\tabletypesize{\footnotesize}
\tablecaption{EXors list.\label{exor:tab}}
\tablewidth{0pt}
\tablehead{
\colhead{Target} & \colhead{$\alpha$ (2000.0) }&
\colhead{$\delta$ (2000.0)} &\colhead{Location} &
\colhead{binary}
}
\startdata
XZ Tau      & 04 31 40.07 & + 18 13 57.1 & L1551  &  Y (0.30) \\
UZ Tau E    & 04 32 42.96 & + 25 52 31.1 & B19    &  Y (0.34) \\
VY Tau      & 04 39 17.42 & + 22 47 53.4 & L1536  &  Y (0.66) \\
DR Tau      & 04 47 06.22 & + 16 58 42.9 & L1558  &  N \\
V1118 Ori   & 05 34 44.66 & - 05 33 41.3 & ONC    &  - \\
NY Ori      & 05 35 35.79 & - 05 12 20.7 & ONC    &  - \\
V1143 Ori   & 05 38 03.89 & - 04 16 42.8 & L1640  &  - \\
EX Lup      & 16 03 05.49 & - 40 18 25.3 & Lup 3  &  N \\
PV Cep      & 20 45 53.96 & + 67 57 38.9 & L1158  &  - \\
%           &       &       &      &    &      &    &  - \\
%\hline\\
\enddata
\end{deluxetable}

In Table~\ref{exor:tab} the list of the EXor variables known so
far is given; the source XZ Tau has been added to the original
Herbig's list (1989) since it has been subsequently recognized as
a potential candidate (Coffey, Downes \& Ray 2004). Indeed, it is
a binary system (Close et al. 1997) where the A (SE) component is
separated by 0.3 arcsec from the B (NW) component.

In the same table is also given indication of the dark cloud to
which the source belongs and whether or not a close companion has
been found, providing in the affirmative cases the inter-distance in
arcsec. The flux ratio in the IR regime between primary and
secondary component has been recently measured (McCabe et al. 2006)
only for VY Tau in the L band, and it equals to 6.1. The first seven
columns of Table~\ref{parameters:tab} summarize the relevant
parameters (distance, visual extinction, luminosity, spectral type,
range of visual magnitude variation) associated to any individual
source; in the eighth column, the occurrence (or not) of mass loss
phenomena in form of both jet and molecular outflow is pointed out.

%%%%   TABLE - OBJECT PARAMETERS  %%%%%%%%%%%%%%%%%%%%%%%%%%%%%%%%%%%%%%%%%%%%%%%%

\begin{deluxetable}{ccccccccc}
\tabletypesize{\footnotesize}
\tablecaption{EXOrs observed parameters.\label{parameters:tab}}
\tablewidth{0pt}
\tablehead{
\colhead{Target} & \colhead{Dist.} &  \colhead{A$_V$} & \colhead{L$_{bol}$}
    &\colhead{SpT} & \colhead{V$_{max}$} & \colhead{V$_{min}$} &
    \colhead{jet/outflow} & \colhead{Ref.}
    }
\startdata
       & (pc)  & (mag)  & ($L_{\sun}$) &     & \multicolumn{2}{c}{(mag)} &  &  \\
\tableline
XZ Tau      &  140 &  3: &4.2 - 10.7& M3 & 10.4 & 16.6   &  Y  & 1,2,3,4,5  \\
UZ Tau E    &  140 & 1.49&   1.7    & M1,3& 11.7  & 15.0 &     & 1,3        \\
VY Tau      &  140 & 0.85&   0.75   & M0V & 9.0   & 15.3 &     & 1,3,6      \\
DR Tau      &  140 &1.7-2.1&1.05-5.0&K7-M0& 10.5  & 16.0 &     & 1,3,7,8    \\
V1118 Ori   &  140 & 0-2   &1.4-25.4& M1e & 14.2  & 17.5 &  N  & 1,9        \\
NY Ori      &  400 & 0.3 &            & A0V & 14.5& 17.5 &     & 1,10       \\
V1143 Ori   &      &     &            & M   &  13 &  19  &     & 1          \\
EX Lup      &  140 &     &   0.47     & M0.5& 8.4 & 14.3 &     & 1,11,12    \\
PV Cep      &  500 &0.4-5&   100      & A5e & 11.1& 18.0 &  Y  & 1,13,14    \\
%           &      &     &            &     &     &      &     &            \\
\enddata

\tablecomments{References to the Table: (1) SIMBAD Astronomical
Database (http://simbad.u-strasbg.fr/simbad); (2) Coffey, Downes
\& Ray 2004; (3) Herbig \& Bell 1988; (4) Carr 1990; (5) Evans II
et al. 1987; (6) Herbig 1990; (7) Cohen \& Kuhi 1979; (8) Kenyon
et al. 1994; (9) Paper I; (10) Breger, Gherz \& Hackwell 1981;
(11) Gras-Vel\'{a}zquez \& Ray 2005; (12) Herbig et al. 2001; (13)
Cohen et al. 1981; (14) Van Citters \& Smith 1989.}
\end{deluxetable}

\section{IR catalog of EXors}

In spite of the importance of the 1-10 $\mu$m spectral region for
understanding the disk accretion processes, little information is
available about the steadiness of the IR emission from the EXor. In
this section all the literature data are compiled in form of a
catalogue. The catalogue includes all the IR (1-10 $\mu$m)
observations known to us by December 2006. In Table~\ref{nir:tab},
for each EXor are given: (i) the date of the observation; (ii) the
results of the JHKLM photometry, where the errors (when available in
the original papers) are expressed in units of 0.01 mag in
parentheses, otherwise they can be assumed in the range
0.1-0.2 mag (JHK) and 0.2-0.3 mag (LM); (iii) the reference of the
observations, where the reader should be able to find the necessary
information for flux calibration, photometric system and effective
wavelengths. Table~\ref{mir:tab} is organized in the same way as
Table~\ref{nir:tab}, but reports the photometry in the 8-13 $\mu$m
range (N band), given as flux density (in Jansky) instead of
magnitudes. Such a spectral range has been arbitrarily divided into
five channels (each one 1 $\mu$m wide), aiming to preserve the
differences which result from narrow band observations carried out
with different filters; however, when a single flux value
corresponds to a given date, it is obtained through wide band
photometry, whose effective wavelength usually falls in the (10-11)
$\mu$m channel. The tables for converting magnitudes into flux
densities are given in the referenced papers. Table~\ref{fir:tab}
lists the far-IR (12-100 $\mu$m) photometry obtained by IRAS/MSX and
mention is given to the ISO (Infrared Space Observatory) detections.
Finally, Table~\ref{spectra:tab} provides the indication and
reference to the available near-,
mid-IR spectroscopy or to the investigation of specific spectral features.\\

As explicitly indicated in the original papers, the observations
have been obtained with a large variety of focal plane
instrumentation. For the purpose of inter-comparing these data, one
has to be careful to not introduce spurious effects, which could
arise from the different photometric systems and/or from having
merged in the same catalogue the more recent data, obtained with IR
arrays, along with the older ones, obtained with
single-detector photometers equipped with larger focal plane
apertures.
The catalogued data may be useful to discriminate stages
of relative activity vs. inactiveness of the source, but, because of
the problems discussed above, they are not directly suitable for
constructing reliable light curves. Molinari, Liseau \& Lorenzetti
(1993) provide a quantitative discussion about the data
inhomogeneity introduced by the difference of the various
calibration systems used during the early IR observations.\\

Some comments on the catalogued data: the 9 EXors have been
photometrically observed in the 1-10 $\mu$m range on 88 occasions in
total, apart of a daily monitoring of DR Tau (44 observations by
Kenyon et al. 1994). Three objects (30 \%) have been observed just 3
times in 30 years. A (quasi-)simultaneous entire SED (1-10 $\mu$m)
has never been obtained, but in six occasions a partial coverage
(typically KLMN bands) has been gathered. The K-band photometry is
the one most frequently observed (about 70 \% of the cases),
but individual sources have a relative low number (from 2 to 11) of
K magnitude determination.\\

%%%%   TABLE - NEAR INFRARED %%%%%%%%%%%%%%%%%%%%%%%%%%%%%%%%%%%%%%%%%%%%%%%%

\begin{deluxetable}{llcccccc}
\tabletypesize{\footnotesize}
\tablecaption{EXor catalog: Near-infrared observations\label{nir:tab}}
\tablewidth{0pt}
\tablehead{
\colhead{Target} & \colhead{Date} &  \colhead{J} & \colhead{H} &
\colhead{K} & \colhead{L} & \colhead{M} & \colhead{ref}
}
\startdata
{\bf XZ Tau$^a$}& Dec 1972 - Apr 1974&    -     &  -       & 7.3      &  5.45       &  4.2    &  1     \\
            & 1973 Oct 5         & 10.00    & 9.29     & 8.57     &  6.18       &  -      &  2     \\
            & 1973 Nov 11        &    -     &  -       & 8.47     &  6.72       &  5.4    &  2     \\
            & 1973 Dec 11        &    -     &  -       & 8.65     &  7.01       &  5.4    &  2     \\
            & 1974 Oct 19        & 10.13    & 9.12     & 8.12     &  6.30       &  -      &  2     \\
            & 1974 Nov 23        & 9.76     & 8.89     & 7.95     &  6.28       &  -      &  2     \\
            & 1981 Dec 3         & 9.34 (2) & 8.16 (1) & 7.17 (1) &  5.85 (4)   &  -      &  3     \\
            & 1986 Dec 7         &    -     &  -       & 7.2      &      -      &  -      &  4     \\
            & 1993 Oct 2         & 9.69     & 8.77     & 7.97     &      -      &  -      &  5     \\
            & 1993 Dec 8-13      &    -     &  -       & 7.43     &      -      &  -      &  5     \\
            & 1999 Nov 11        & 9.38 (2) & 8.15 (4) & 7.29 (2) &      -      &  -      &  6     \\
\tableline
{\bf UZ Tau E}   & Dec 1972 - Apr 1974 &    -     &  -       & 7.3      &  6.2  &  -      &  1     \\
           & 1976 Nov 1          & 8.72     & 7.83     & 7.26     &  6.29       &  -      &  7     \\
           & 1976 Dec 18         & 8.45     & 7.60     & 7.02     &  6.40       &  -      &  7     \\
           & 1977 Nov 4          &    -     & 7.97 (1) & 7.14 (1) &  6.04 (2)   &  -      &  8     \\
           & 1981 Jan 12         &    -     &  -       & 7.34     &  6.4 (10)   & 5.6 (30)&  9     \\
           & 1986 Nov 11         &    -     &  -       & 7.0      &      -      &  -      &  4     \\
           & 1988 Jan 15         & 9.83 (2) & 8.46 (2) & 7.59 (2) &      -      &  -      &  10    \\
           & 1990 Oct 8          &    -     &  -       & 7.3      &      -      &  -      &  11    \\
           & 1997 Nov 30         & 9.14 (4) & 8.12 (1) & 7.35 (3) &      -      &  -      &  6     \\
           & 2001 Dec 1          &    -     &  -       & 7.60 (10)&  6.78 (5)   &  -      &  28    \\
           & 2004 Mar 6          &    -     &  -       &   -      &  5.98 (3)   &5.46 (4) &  12    \\
\tableline
{\bf VY Tau}   & Dec 1972 - Apr 1974 &    -     &  -       & 8.65 &  8.6        &$>$ 5.2  &  1     \\
           & 1973 Oct 9          & 9.71     & 9.22     & 9.16     &      -      &  -      &  2     \\
           & 1973 Dec 13         &    -     &  -       & 8.94     &  8.34       &  -      &  2     \\
           & 1977 Nov 4          &    -     & 9.51 (1) & 9.03 (2) &  8.87 (13)  &  -      &  8     \\
           & 1981 Dec 2          & 10.03 (2)& 9.28 (2) & 8.97 (2) &  8.50 (6)   &  -      &  3     \\
           & 1990 Oct 5          & 10.13    & 9.40     & 9.08     &      -      &  -      &  13    \\
           & 1993 - 1998         &10.15 (29)& 9.50 (1) & 9.22 (18)&      -      &  -      &  14    \\
           & 1998 Oct 3          & 9.97 (2) & 9.26 (2) & 8.96 (2) &      -      &  -      &  6     \\
           & 2001 Dec 1          &    -     &  -       &   -      &  8.31 (9)   &  -      &  28    \\
%\hline\\[-5pt]
           &                     &          &          &          &             &         &        \\
\tableline
{\bf DR Tau}     & Dec 1972 - Apr 1974 &    -     &  -       & 7.2      &  5.9        &  -&  1     \\
           & 1973 Nov 12         &    -     &  -       & 7.47     &  5.85       & 5.0 (20)&  2     \\
           & 1977 Nov 4          &    -     & 7.54 (1) & 6.21 (2) &  4.83 (5)   &  -      &  8     \\
           & 1981 Dec 2          & 8.79 (2)& 7.66 (1) & 6.65 (1)  &  5.34 (4)   &  -      &  3     \\
           & 1981 Dec 3          & 8.54 (1)& 7.46 (1) & 6.45 (1)  &  5.15 (4)   &  -      &  3     \\
           & 1981 Dec 16         &    -    &   -      &   -       &      -      &  4.17   &  9     \\
           & 1986 Dec 5          &    -    &   -      &  6.9      &      -      &  -      &  4     \\
           & 1987 Nov (7-24)     &8.58-9.07&7.48-7.92 &6.48-6.91  &5.01-5.32    &4.47-4.69&  15 - 20 obs.  \\
           & 1988 Sep (1-13)     &8.44-9.24&7.40-8.06 &6.41-7.04  &4.93-5.46    &  -      &  15 - 24 obs.  \\
           & 1996-1998           &         &          &           & 5.4         &  -      &  16     \\
           & 1997 Oct 10         & 8.84 (2)& 7.80 (5) & 6.87 (1)  &      -      &  -      &  6     \\
           & 1998 May - 1999 Jan &8.79-9.19&   -      &6.76-7.12  &      -      &  -      &  17 - 4 obs.   \\
           & 2004 Mar 5          &    -     &  -      &   -       &      -      &4.50 (1) &  12     \\
\tableline
{\bf V1118 Ori}  & 1991-1993    &   12.21 (1)  & 11.29 (1)  &  10.49 (1) &  -     &    -   &  18   \\
           & 1994 Mar 15  &   8.52 (5)   & 8.04 (4)   &  7.94 (3)  &  -     &    -   &  19   \\
           & 2000 Nov 24  &   12.64 (2)  & 11.51 (3)  &  10.85 (2) &  -     &    -   &  6    \\
           & 2005 Mar 20  &   11.16 (2)  & -          &  9.79 (1)  &  -     &    -   &  20   \\
           & 2005 Apr 3   &   10.79 (1)  & -          &  9.48 (1)  &  -     &    -   &  20   \\
           & 2005 Apr 15  &   10.94 (2)  & -          &  9.53 (1)  &  -     &    -   &  20   \\
           & 2005 Sep 6   &   11.08 (1)  & -          &  9.71 (2)  &  -     &    -   &  20   \\
           & 2005 Sep 11  &   11.23 (2)  & 10.45 (1)  &  9.85 (1)  &  -     &    -   &  20   \\
\tableline
{\bf NY Ori}     & 1975 Sep 22-29      &    -     & 9.7      & 9.64     &  $\geq$ 9.0 &  -&   21  \\
           & 2000 Nov 24         & 9.75 (4) & 9.74 (3) & 9.73 (3) &      -      &  -      &   6   \\
\tableline
{\bf V1143 Ori}  & 1994 Mar 15         &12.51 (17)&11.79 (12)&11.41 (12)&        &        &   22  \\
           & 1998 Oct 30         &12.63 (2) &11.96 (2) &11.58 (2) &      -      &  -      &   6   \\
\tableline
{\bf EX Lup}     & 1972 Jun 5    & 9.76 (13)& 9.04 (14)&8.82 (12) & $>$ 8.7     &  -&   23  \\
           & 1982 Apr            & 9.92 (1) & 9.11 (1) & 8.78 (1) & 8.05 (1)    &7.54 (5) &   24  \\
           & 1999 May 16         & 9.73 (2) & 8.96 (2) & 8.50 (2) &      -      &  -      &   6   \\
\tableline
{\bf PV Cep}     & $<$ 1982            & 10.57    & 8.60     & 6.75     &      -      &  -      &   25  \\
           & $<$ 1984            & 10.39    & 8.32     & 6.76     & 4.63        &3.76     &   26  \\
           & 1992 Aug 6-8        & 9.39 (1) & 7.88 (2) & 6.44 (3) &      -      &  -      &   27  \\
           & 1999 Sep 28         &12.45 (2) & 9.50 (3) & 7.29 (1) &      -      &  -      &   6   \\
%\hline\\[-5pt]
           &                     &          &          &          &             &         &      \\
\enddata

\tablecomments{References to the Table: (1) Cohen 1974; (2)
Rydgren, Strom \& Strom 1976; (3) Rydgren \& Vrba 1983; (4) Tamura
\& Sato 1989; (5) Whitney, Kenyon \& Gomez 1997; (6) 2MASS - Cutri
et al. 2003; (7) Elias 1978; (8) Cohen \& Kuhi 1979; (9) Rydgren
et al. 1984; (10) Moneti \& Zinnecker 1991; (11) Simon et al.
1992; (12) Hartmann et al. 2005; (13) Kenyon \& Hartmann 1995;
(14) Woitas, Leinert \& K\"{o}hler 2001; (15) Kenyon et al. 1994;
(16) Thi et al. 2001; (17) Eiroa et al. 2002; (18) Hillenbrand et
al. 1998; (19) Garc\'{i}a et al. 1995; (20) Paper I;
(21) Breger, Gehrz \& Hackwell1 1981; (22) Mampaso \& Parsamian
1995; (23) Glass \& Penston 1974; (24) Appenzeller, Jankovics \&
Krautter 1983; (25) Lacasse 1982; (26) Neckel \& Staude 1984; (27) Li et al. 1994; (28) McCabe et al. 2006.}
\tablenotetext{a}{It is a binary system (see text), whose A
component is slightly brighter in J band but fainter in K. As a
consequence, near-IR data reasonably account for both sources and
thus should be treated with enough care.}
\end{deluxetable}
\clearpage

%%%%   TABLE - MID-INFRARED %%%%%%%%%%%%%%%%%%%%%%%%%%%%%%%%%%%%%%%%%%%%%%%%
\begin{deluxetable}{llcccccc}
\tabletypesize{\footnotesize}
\tablecaption{ EXor catalog: N band (8-13 $\mu$m) ground-based observations\label{mir:tab}}
\tablewidth{0pt}
\tablehead{
\colhead{Target} & \colhead{Date} & \colhead{F(8-9)$^a$}& \colhead{F(9-10)} &
\colhead{F(10-11)} & \colhead{F(11-12)} & \colhead{F(12-13)} & \colhead{ref}
}
\startdata
{\bf XZ Tau}     & Dec 1972 - Apr 1974&    -     &  -       & 6.3       &      -      &   -     &  1    \\
            & 1973 Nov 11        &  1.9     &  -       &   -       &  2.8        &  2.4         &  2    \\
            & 1973 Dec 11        &  1.5     &  -       &   -       &  1.6        &  2.0         &  2    \\
            & 1999 Nov 17        &    -     &  -       &  4.61     &   -         &  -           &  10   \\
\tableline
{\bf UZ Tau E}   & Dec 1972 - Apr 1974 &    -     &  -       &  1.6      &      -      &  -     &  1    \\
           & 1976 Nov 1          &    -     &  -       &  1.3      &      -      &  -           &  3    \\
           & 1976 Dec 18         &    -     &  -       &  1.0      &      -      &  -           &  3    \\
           & 1981 Jan 12         &  1.0     &  -       &  1.1      &  1.1        &  0.9         &  4    \\
           & 1992 Dec 8          &  0.74    &   1.06   &   -       &      -      &  0.84        &  5    \\
           & 1999 Nov 16         &    -     &  -       &  1.35     &   -         &  -           &  10   \\
           & 2004 Mar 6          &  1.2     &  -       &   -       &      -      &  -           &  6    \\
\tableline
{\bf VY Tau}& 1999 Nov 17        &    -     &  -       &  1.17     &   -         &  -           &  10   \\
\tableline
{\bf DR Tau}    & Dec 1972 - Apr 1974 &    -      &  -       &  2.2     &      -      &  -      &  1    \\
           & 1973 Nov 12         &   1.1     &  -       &  -       &  1.3        &  1.5         &  2    \\
           & 1981 Dec 16         &   3.8     &  -       &  3.8     &  3.4        &  3.1         &  4    \\
           & 1982 Dec 20         &    -      &  -       &  2.9     &      -      &  -           &  4    \\
           & 1996-1998           &    -      &  2.4     &   -      &      -      &  -           &  8    \\
           & 2002 Dec 24-27      &    -      &  -       &   -      &  2.0        &  -           &  9   \\
           & 2004 Mar 5          &  1.99     &  -       &   -      &      -      &  -           &  6    \\
\tableline
%{\bf V1118 Ori}  & 2006 Jan 12   &           &    -     &  0.07     &      -      &             &  7    \\
%\hline \\[-5pt]
{\bf PV Cep}  &  1996-1998    &  9.1      &  8.0     &  10.1     & 14.0        &  15.3          &  7    \\
              &               &           &          &           &             &                &       \\
\enddata
\tablecomments{References to the Table: (1) Cohen 1974; (2) Rydgren, Strom \&
Strom, 1976; (3) Elias 1978; (4) Rydgren et al. 1984; (5) Ghez et
al. 1994; (6) Hartmann et al. 2005; (7) SWS: Acke \& van den Ancker
2004;(8) Thi et al. 2001; (9) Przygodda et al. 2003; (10) McCabe et al.
2006.}
\tablenotetext{a}{
F($\lambda_1$-$\lambda_2$) indicates the flux (in Jansky)
obtained at an effective wavelength in the range
$\lambda_1$-$\lambda_2$ $\mu$m.}
\tablenotetext{-}{Errors are always lesser than 10\%.}
\tablenotetext{-}{Zero-magnitude fluxes for converting mag into Jy, are
given by the authors. Otherwise, the adopted conversion table is
given in
$http://www.jach.hawaii.edu/UKIRT/astronomy/utils/conver.html$.\\
}
\end{deluxetable}

%%%%   TABLE - FAR-INFRARED %%%%%%%%%%%%%%%%%%%%%%%%%%%%%%%%%%%%%%%%%%%%%%%%
\footnotesize
\begin{table}
\begin{center}

\caption{EXor catalog: Far-IR observations\label{fir:tab}}
\begin{tabular}{lcccccccc}
\tableline\tableline
Target\tablenotemark{a} &\multicolumn{4}{c}{IRAS\tablenotemark{b}} & ref
& \multicolumn{3}{c}{ISO\tablenotemark{c}}\\
\tableline
       & 12$\mu$m & 25$\mu$m & 60$\mu$m & 100$\mu$m &   & SWS & LWS & PHT  \\
\tableline
{\bf UZ Tau E}  & 1.38 (0.03) & 1.76 (0.04)  & 2.37 (0.07)  & 1.2 (0.6)  & 1 & - & - & -   \\
{\bf VY Tau}    & 0.20 (0.03) & 0.29 (0.03)  & 0.34 (0.04)  & 1.2 (0.3)  & 1 & - & - & -   \\
{\bf DR Tau}    & 3.16 (0.03) & 4.30 (0.05)  & 5.51 (0.04)  & 7 (1)      & 1 & 2 & - & -   \\
{\bf NY Ori}    & 10 (3)      &       -      &      -       &     -      & 1 & - & - & -   \\
{\bf V1143 Ori} & 0.17 (0.04) & 0.10 (0.06)  & 2.7 (0.3)    & 9 (3)      & 1 & - & - & -   \\
{\bf EX Lup}    & 0.71 (0.05) & 1.09 (0.03)  & 1.25 (0.05)  & 3 (2)      & 1 & - & - & -   \\
{\bf PV Cep}    & 13.27 (0.03)& 33.18 (0.04) & 51.87 (0.06) & 51.2 (0.2) & 1 & 3 & 4 & 3,5 \\
                & 13.4$^d$    & 23.4$^d$     & 40.3$^d$     & 51.0$^d$   & 6 &   &   &     \\
\tableline
\end{tabular}

\tablecomments{References to the Table: (1) Weaver \& Jones 1992; (2) Thi et al.
2001; (3) Acke \& van den Ancker 2004;
(4) Lorenzetti et al. 1999; (5) \'{A}brah\'{a}m et al. 2000; (6) Elia et al. 2005.}

\tablenotetext{a}{MSX data of the source V1118 Ori are given in
Table~\ref{Jy:tab}.}
\tablenotetext{b}{Co-added flux densities and errors are given in Jy.}
\tablenotetext{c}{On board of ISO satellite: SWS (Short Wavelength
Spectrometer - 3-45$\mu$m); LWS (Long Wavelength Spectrometer
45-200$\mu$m); PHT (Imager with spectroscopic capability
3-200$\mu$m).}
\tablenotetext{d}{Values derived by ISO spectrophotometry convolved with
IRAS bandpasses.}
\end{center}
\end{table}

%%%%   TABLE - IR SPECTROSCOPY  %%%%%%%%%%%%%%%%%%%%%%%%%%%%%%%%%%%%%%%%%%%%%%%%

\begin{table}
\begin{center}
\caption{EXOrs IR spectroscopy data.\label{spectra:tab}}
\begin{tabular}{lc}
\tableline\tableline
Target &      IR spectroscopic features            \\
\tableline
XZ Tau      &  near- (1); (8-13 $\mu$m) sil.? (2); Pa$\beta$,
Br$\gamma$ (3); Br$\gamma$, Br$\alpha$ (4); (5-35 $\mu$m) sil.? (22) \\
            & 2-4$\mu$m (5); Br$\gamma$, H$_2$ (6); Pa$\beta$, Br$\gamma$ (7); CO bands abs.(8); HeI, Pa$\gamma$ (19) \\
UZ Tau E    &   (8-13 $\mu$m) sil.em. (2,23); near- (9); HeI, Pa$\gamma$ (19); (5-35 $\mu$m) sil.em. (22)      \\
VY Tau      &   Pa$\beta$ (3); near- (1),(10); HeI, Pa$\gamma$ (19); (5-35 $\mu$m) sil.em. (22)             \\
DR Tau      &   near-(1); Pa$\beta$, Br$\gamma$ (3); mid-(2); HeI (11);  Pa$\beta$, Br$\gamma$ (7),(12);    \\
            &   H$_2$ rot. - upp.lim. (13); (8-13 $\mu$m) sil.em. (20); (5-35 $\mu$m) sil.em. (22)          \\
V1118 Ori   &   near-, CO bands em. (14)                                                                    \\
NY Ori      &                                                                                               \\
V1143 Ori   &                                                                                               \\
EX Lup      &   near- (15); (5-35 $\mu$m) sil.em. (21)                                                      \\
PV Cep      &   Br$\gamma$, H$_2$ (6); CO (8); SWS: 2-15$\mu$m (16); [FeII] (17); 2-4$\mu$m (5)(18)         \\
%           &      &     &            &     &     &      &  &                                               \\
\tableline
\end{tabular}
\tablecomments{References to the Table: (1) Folha \& Emerson 1999; (2) Cohen \&
Witteborn 1985; (3) Folha \& Emerson 2001; (4) Evans et al. 1987;
(5) Sato et al. 1990; (6) Carr 1990; (7) Giovanardi et al. 1991; (8)
Biscaya et al. 1997; (9) Elias 1978; (10) Greene \& Lada 1996; (11)
Edwards et al. 2003; (12) Muzerolle, Hartmann \& Calvet (1998); (13)
Thi et al. 2001; (14) Paper I; (15) Herbig et al. 2001; (16) Acke \&
van den Ancker 2004; (17) Hamann et al. 1994; (18) van Citters \&
Smith 1989; (19) Edwards et al. 2006; (20) Przygodda et al. 2003;
(21) Kessler-Silacci et al. 2006; (22) Furlan et al. 2006; (23) Honda et al. 2006.
}
\end{center}
\end{table}

\normalsize
\clearpage

\section{New observations of V1118 Ori}

We present a set of new observations of V1118 Ori carried out in the
optical and IR bands during the declining phase (from Oct. 2005
to Dec. 2006) that followed the last outburst occurred at the end
of 2004. Complementary data that sample both the outburst phase and
the beginning of the declining in the near-IR and X-ray regimes are
given in Audard et al. (2005) and in Paper I; this latter provides
also information since the first documented burst during the period
1982-1984.

\subsection{I band polarimetry}

The polarimetric observations were obtained in the I band during
the period November 2005 - December 2006, with the
photometer-polarimeter of the Astronomical Institute of
St.Petersburg University mounted on the 70 cm telescope of the
Crimean Observatory (Ukraine). This photometer is based on ST-7XME
SBIG CCD and provides a polarimetric mode by means of two Savart
plates, rotated by 45$^{\circ}$ one relative to another and used
as analyzer, giving $q$ and $u$ Stokes parameters. The pixel scale
is 0.64 arcsec/pxl which corresponds to a 8.1$\times$5.4
arcmin$^2$ field of view. Within this field the star located 80
arcsec to the North of V1118 Ori was also sampled as a reference.
With the same equipment the BVRI photometric data were obtained as
well; they confirm in the four visual bands the V-band results
presented by Garc\'{i}a, Parsamian \& Velazquez (2006) and are not
presented in this paper which is more oriented to the IR
properties of EXors.

\subsection{Near-IR photometry}

\begin{figure}
 \centering
   \plotone{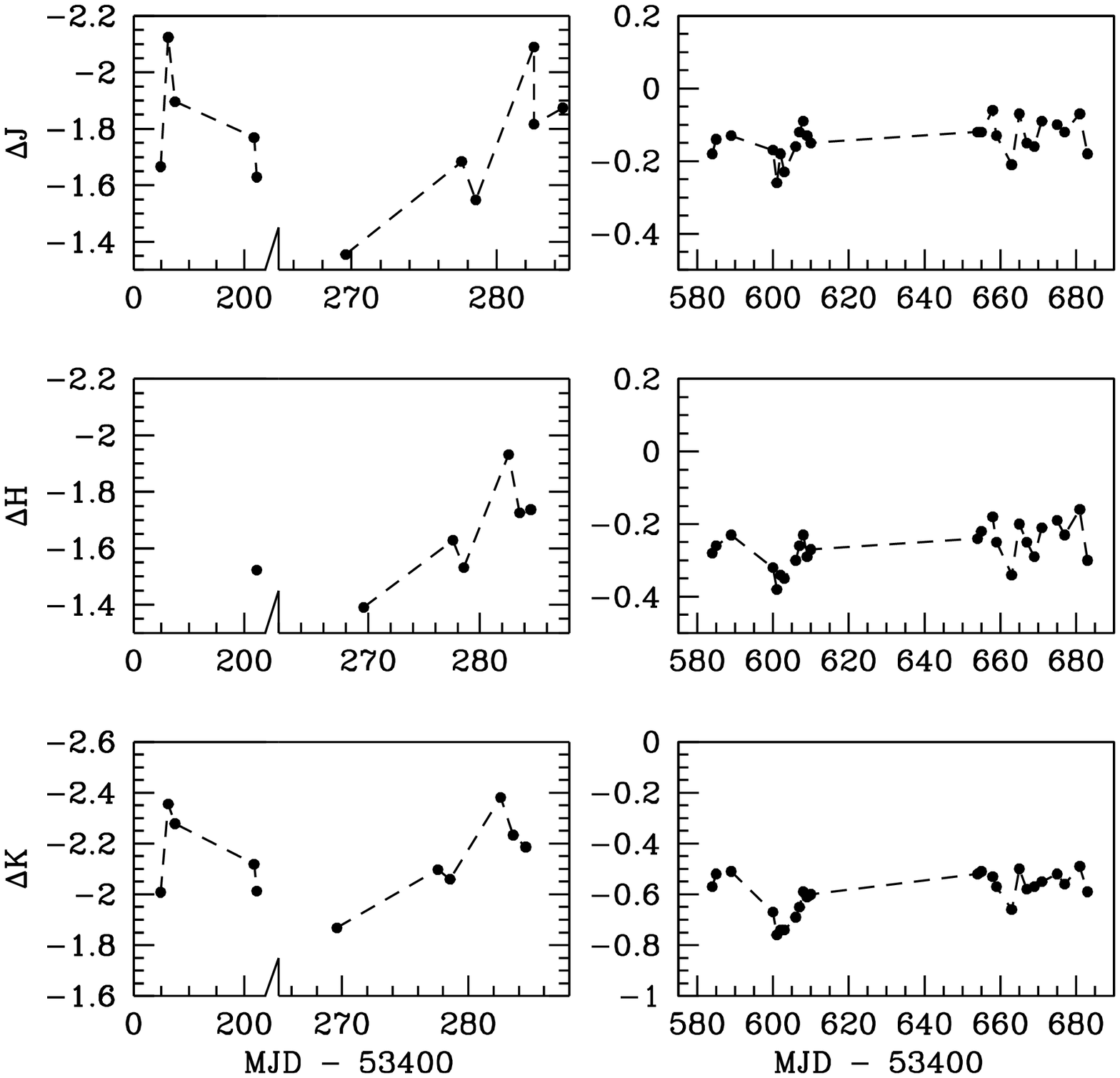}
   \caption{V1118 Ori near-IR light curves vs. MJD (Modified Julian Date). Magnitudes are
   relative to a comparison field star (see text, $\Delta$ = mag$_{obj}$ - mag$_{ref}$). The left panels refer
   to the J,H and K bands observations carried out during two monitoring
   periods. Data points of the first period (Mar-Sept 2005) are
   given in Paper I, while those of the second period (Oct-Nov 2005)
   are listed in Table~\ref{mag:tab}. Here two different
   scales occur in the x-axis, that refer to monitoring periods of different
   duration. Right panels depict the 2006 observations, when the
   quiescence phase had been already reached.
   \label{lightcurve:fig}}
\end{figure}

Near-IR data were obtained at the 1.1 m AZT-24 telescope located at
Campo Imperatore (L'Aquila - Italy) equipped with the
imager/spectrometer SWIRCAM (D'Alessio et al. 2000), which is based
on a 256$\times$256 HgCdTe PICNIC array. Photometry is performed
with broad band filters J (1.25 $\mu$m), H (1.65 $\mu$m), and K
(2.20 $\mu$m). The total field of view is 4.4$\times$4.4 arcmin$^2$,
which corresponds to a plate scale of 1.04 arcsec/pixel. All the
observations were obtained by dithering the telescope around the
pointed position. The raw imaging data were reduced by using
standard procedures for bad pixel removal, flat fielding, and sky
subtraction. The photometric data are presented in
Table~\ref{mag:tab}. Several comparison stars are available within
the 4$\arcmin$.4 SWIRCAM field; their magnitudes result stable
within 0.01 mag in JHK during the period of our observations; in
particular one of them is used to construct the near IR light curves
in terms of differential photometry given in
Figure~\ref{lightcurve:fig}.

%%%%%%%%%%%%%%%%%%%%%%%%%%%%%%%  FIGURE  LIGHTCURVE %%%%%%%%%%%%%%%%%%%%%%%%%%%%%%%%%%%%%%%%%%%%
%%%%%%%%%%%%%%%%%%%%%%%%%%%%%%%%%%%%%%%%%%%%%%%%%%%%%%%%%%%%%%%%%%%%%%%%%%%%%%%%%%%%%%%%%%%%%%%%%%%%%

%%%%   TABLE - MAG  %%%%%%%%%%%%%%%%%%%%%%%%%%%%%%%%%%%%%%%%%%%%%%%%
\begin{deluxetable}{ccccccc}
\tabletypesize{\scriptsize} \tablecaption{Near-IR photometry of
V1118 Ori. A horizontal line separates the 2005 observations from
the 2006 ones. \label{mag:tab}} \tablewidth{0pt} \tablehead{
\colhead{UT date}  &  \colhead{MJD$^a$}  & \colhead{J$^b$} &
\colhead{H} & \colhead{K} & \colhead{[J-H]} & \colhead{[H-K]} }
\startdata
05 Oct 26    &  53669  &  11.46    & 10.52   &  9.92  & 0.94  &  0.60   \\
05 Nov 03    &  53677  &  11.16    & 10.34   &  9.71  & 0.82  &  0.63   \\
05 Nov 04    &  53678  &  11.27    & 10.41   &  9.74  & 0.86  &  0.67   \\
05 Nov 08    &  53682  &  10.78    & 10.04   &  9.47  & 0.74  &  0.57   \\
05 Nov 09    &  53683  &  11.03    & 10.23   &  9.61  & 0.80  &  0.62   \\
05 Nov 10    &  53684  &  10.96    & 10.18   &  9.60  & 0.78  &  0.58   \\
\tableline
06 Sep 06    &  53984  &  12.63    & 11.61   & 10.97  & 1.02  &  0.64   \\
06 Sep 07    &  53985  &  12.62    & 11.62   & 10.97  & 1.00  &  0.65   \\
06 Sep 11    &  53989  &  12.66    & 11.66   & 10.98  & 1.00  &  0.68   \\
06 Sep 22    &  54000  &  12.63    & 11.57   & 10.82  & 1.06  &  0.75   \\
06 Sep 23    &  54001  &  12.54    & 11.51   & 10.74  & 1.03  &  0.77   \\
06 Sep 24    &  54002  &  12.63    & 11.57   & 10.78  & 1.06  &  0.79   \\
06 Sep 25    &  54003  &  12.59    & 11.55   & 10.78  & 1.04  &  0.77   \\
06 Sep 27    &  54006  &  12.62    & 11.60   & 10.85  & 1.02  &  0.75   \\
06 Sep 28    &  54007  &  12.67    & 11.63   & 10.89  & 1.04  &  0.74   \\
06 Sep 29    &  54008  &  12.66    & 11.65   & 10.92  & 1.01  &  0.73   \\
06 Sep 30    &  54009  &  12.65    & 11.60   & 10.91  & 1.05  &  0.69   \\
06 Oct 01    &  54010  &  12.63    & 11.61   & 10.91  & 1.02  &  0.70   \\
06 Nov 15    &  54054  &  12.66    & 11.67   & 11.03  & 0.99  &  0.64   \\
06 Nov 16    &  54055  &  12.69    & 11.69   & 11.06  & 1.00  &  0.63   \\
06 Nov 19    &  54058  &  12.66    & 11.64   & 10.97  & 1.02  &  0.67   \\
06 Nov 20    &  54059  &  12.69    & 11.65   & 10.98  & 1.04  &  0.67   \\
06 Nov 24    &  54063  &  12.57    & 11.54   & 10.91  & 1.03  &  0.63   \\
06 Nov 26    &  54065  &  12.74    & 11.72   & 11.04  & 1.03  &  0.68   \\
06 Nov 28    &  54067  &  12.60    & 11.61   & 10.92  & 0.99  &  0.69   \\
06 Nov 30    &  54069  &  12.61    & 11.62   & 10.94  & 0.99  &  0.68   \\
06 Dec 02    &  54071  &  12.69    & 11.67   & 10.97  & 1.02  &  0.70   \\
06 Dec 06    &  54075  &  12.69    & 11.69   & 11.00  & 1.00  &  0.69   \\
06 Dec 08    &  54077  &  12.63    & 11.64   & 10.93  & 0.99  &  0.71   \\
06 Dec 12    &  54081  &  12.70    & 11.68   & 11.01  & 1.02  &  0.67   \\
06 Dec 14    &  54083  &  12.60    & 11.59   & 10.92  & 1.01  &  0.67   \\
\enddata
\tablenotetext{a}{MJD = modified Julian Date.}
\tablenotetext{b}{Errors of our photometry in all the three bands never
exceed 0.02 mag.}
\end{deluxetable}

%%%%%%%%%%%%%%%%%%%%%%%%%%%%%%%%%%%%%%%%%%%%%%%%%%%%%%%%%%%%%%%%%%%%%%%%

\subsection{Near-IR spectroscopy}

Low resolution ($\mathcal{R}$ $\sim$ 250) spectroscopy has been
obtained with the same instrument (SWIRCAM) described above, by
means of two IR grisms G$_{blue}$ and G$_{red}$ covering the ZJ
(0.83 - 1.34 $\mu$m) and HK (1.44 - 2.35 $\mu$m) bands,
respectively. The long slit is not orientable in position angle, and
it samples a portion of the focal plane, 2$\times$260 arcsec$^2$ in
the east-west direction.

Long-slit spectroscopy was carried out in the standard
ABB$\arcmin$A$\arcmin$ mode with a total integration time of 800
sec. The observations were flat-fielded, sky-subtracted, and
corrected for the optical distortion in both the spatial and
spectral directions. Atmospheric features were removed by
dividing the extracted spectra by that of a normalized telluric
standard star, once corrected for its intrinsic spectral features.
Wavelength calibration was derived from the OH lines present in the
raw spectral images, while flux calibration was obtained from our
photometric data.

\begin{figure}
 \centering
   \plotone{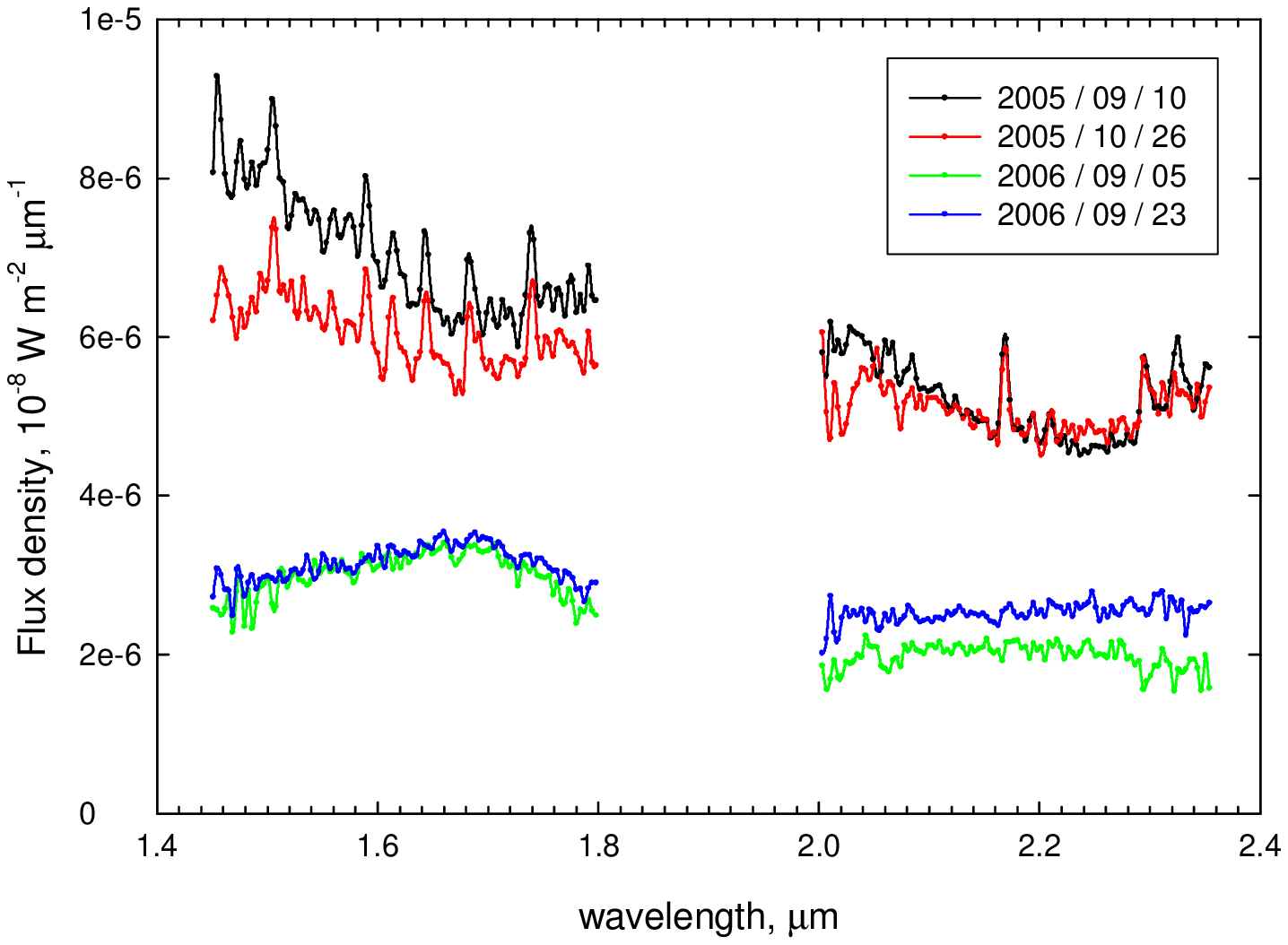}
   \caption{Near-IR spectra (1.44-2.35 $\mu$m) of V1118 Ori taken in different epochs.
Those depicted in the upper part have been carried out in autumn 2005, while those in the lower part were taken a year later, when the star faded by 1.5 mag in the K band.
   \label{nirspectra:fig}}
\end{figure}

We obtained a second near-IR spectrum on 2005 October 26, namely
fifty days later than the first-one (on 2005 September 10, see Paper
I), covering the band from 0.8 to 2.3 $\mu$m, although in
Figure~\ref{nirspectra:fig} is depicted only the HK portion. Although the second spectrum appears more noisy, all the emission
features (from Hydrogen recombination and other ionic and molecular
species) identified in the first spectrum are very well
recognizable and do not show any flux variation within the errors.

However, the real novelty is represented here by a third
spectrum (2006 September 5) taken one year later, covering just the
HK part of the near-IR band and shown in Figure~\ref{nirspectra:fig}
together with the previous spectra. Surprisingly, no emission line
is detected, at our sensitivity, and the continuum shows
a redder shape. At our knowledge, such a dramatic variation was never observed before in an EXor source. Moreover, a confirmation to this featureless appearance is provided by a fourth spectrum, acquired a couple of weeks later and depicted in
Figure~\ref{nirspectra:fig}, as well.

%%%%%%%%%%%%%%%%%%%%%%%%%%%%%%%  FIGURE  NIR SPECTRA %%%%%%%%%%%%%%%%%%%%%%%%%%%%%%%%%%%%%%%%%%%%
%%%%%%%%%%%%%%%%%%%%%%%%%%%%%%%%%%%%%%%%%%%%%%%%%%%%%%%%%%%%%%%%%%%%%%%%%%%%%%%%%%%%%%%%%%%%%%%%%%%%%

\subsection{N-band imaging}

Imaging in the N10.4 broadband filter was carried out on  January
2006 with TIMMI2 (Saviane \& Doublier 2005) at the 3.6m ESO
telescope (La Silla, Chile). The adopted plate scale is 0.3
arcsec/pxl corresponding to a 96$^{\prime\prime} \times$
72$^{\prime\prime}$ field of view. The observations were obtained by
chopping the signal and by nodding and jittering the telescope
around the pointed position in the usual ABB$^{\prime}$A$^{\prime}$
mode. The raw data were reduced by using standard procedures for bad
pixel removal, flat fielding, and sky subtraction. The observed
field was flux calibrated by using the photometric standard star
HD29291. The only two objects present in the field are V1118
Ori and the brighter source V372 Ori, located 47 arcsec in the
SE direction; both present a point-like appearance without any sign
of diffuse emission. The photometric results are given in
Table~\ref{Jy:tab}. In the same Table other observations are
presented: they consist of both {\it Spitzer} photometry at
4.5 and 8.0 $\mu$m, that we have obtained by reducing maps
from that archive and MSX data (Paper I), listed for completeness.
In this respect is noticeable that the {\it Spitzer} IRAC 4.5 $\mu$m
channel contains several H$_2$ lines that are commonly excited
in shocked molecular gas (e.g. Smith 1995). Hence, the lack of any
detectable 4.5 $\mu$m emission in the proximity of V1118 Ori,
provides a further clue to exclude this star as the driver of a molecular jet.

Those presented in Table~\ref{Jy:tab} are the first observations
of V1118 Ori in the mid-IR. Although these data refer to different
effective wavelengths and different epochs, ground-based values
agree with {\it Spitzer} data, but they are largely lower than
IRAS/MSX determinations. These discrepancies have been remarked
several times in the literature concerning YSO's (e.g. Walsh et
al. 2001) and may be due to the larger environmental contamination
suffered by the larger IRAS/MSX beams.

%%%%   TABLE - Jy  %%%%%%%%%%%%%%%%%%%%%%%%%%%%%%%%%%%%%%%%%%%%%%%%
\begin{deluxetable}{ccccc}
\tabletypesize{\footnotesize}
\tablecaption{Mid-IR photometry\label{Jy:tab}}
\tablewidth{0pt}
\tablehead{
\colhead{object} & \colhead{UT date} & \colhead{MJD$^a$} & \colhead{Band} &
\colhead{Flux (Jy)}
}
\startdata
V1118 Ori    & 06 Jan 12    &  53747  & N(10.4 $\mu$m)     & 0.07 $\pm$ 0.01             \\
             & 04 Oct 27    &  53005  & Spitzer 8.0$\mu$m  & (52.70 $\pm$ 0.03) 10$^{-3}$\\
             & 04 Oct 27    &  53005  & Spitzer 4.5$\mu$m  & (42.91 $\pm$ 0.03) 10$^{-3}$\\
             &              &         &                    & (9.05 mag)                  \\
             & 02 Jan 09    &  52283  & MSX 8.28$\mu$m     & 0.3                         \\
             & 02 Jan 09    &  52283  & MSX 12.13$\mu$m    & $<$ 0.5                     \\
\tableline
V372 Ori$^b$ & 06 Jan 12    &  53747  & N(10.4 $\mu$m)     & 1.97 $\pm$ 0.03             \\
             & 02 Jan 09    &  52283  & MSX 8.28$\mu$m     & 2.8                         \\
             & 02 Jan 09    &  52283  & MSX 12.13$\mu$m    & 3.1                         \\
             & -            &    -    & IRAS 12$\mu$m      & 5.0                         \\
\enddata
\tablenotetext{a}{MJD = modified Julian Date.}
\tablenotetext{b}{This source does not belong to our sample, however it is reported
for comparison. It is located about 45 arcsec SE of V1118 Ori.}
\end{deluxetable}
%%%%%%%%%%%%%%%%%%%%%%%%%%%%%%%%%%%%%%%%%%%%%%%%%%%%%%%%%%%%%%%%%%%%%%%%

\section{Results and discussion}

\subsection{Catalogue results}

%%%%%%%%%%%%%%%%%%%%%%%%%%%%%%%  FIGURE JK %%%%%%%%%%%%%%%%%%%%%%%%%%%%%%%%%%%%%%%%%%%%
\begin{figure}
 \centering
   \plotone{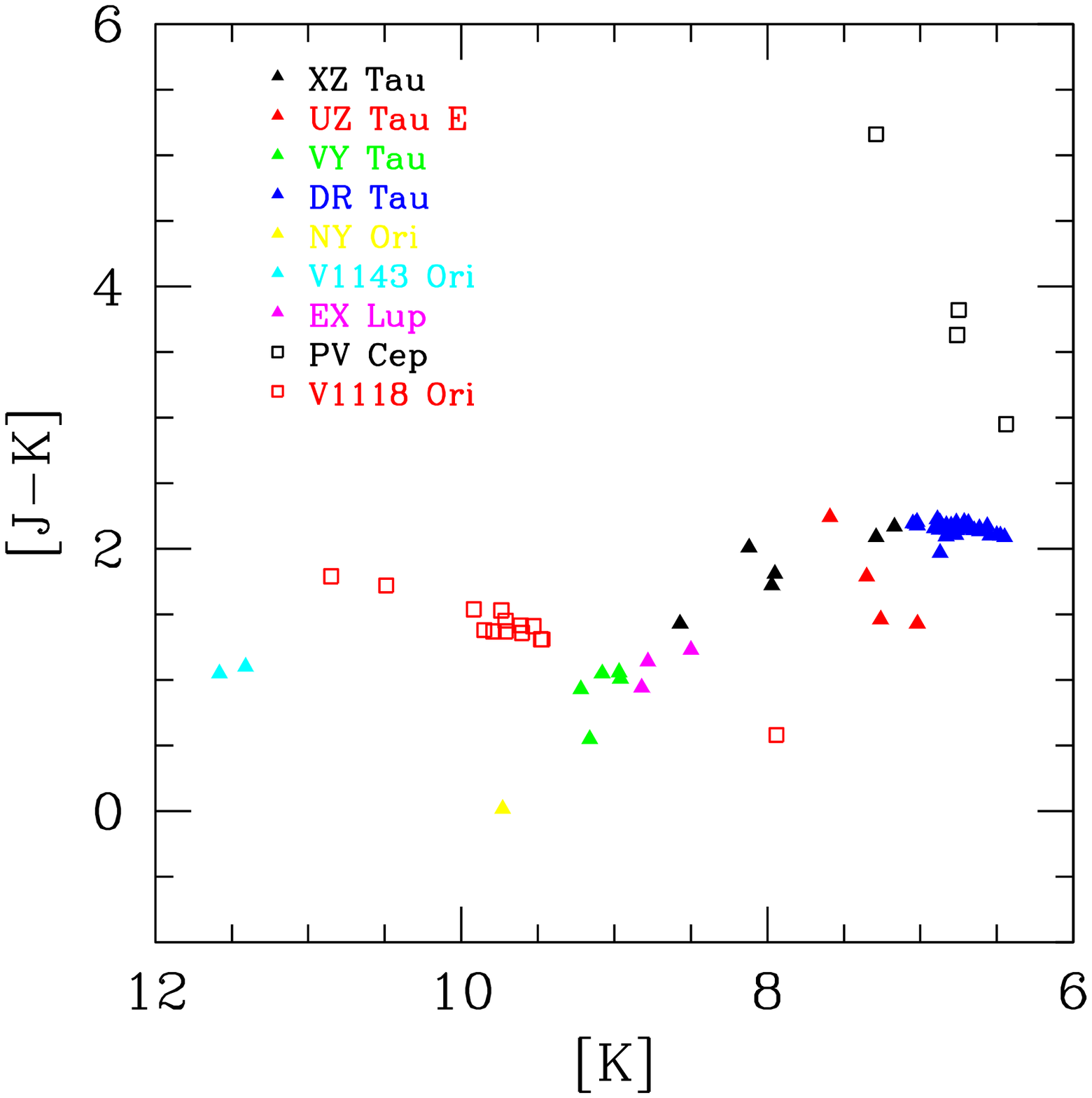}
   \caption{[J-K] colour plotted vs K magnitude for all the EXOr observations listed in Table~\ref{nir:tab}.
   \label{J-K:fig}}
\end{figure}
%%%%%%%%%%%%%%%%%%%%%%%%%%%%%%%%%%%%%%%%%%%%%%%%%%%%%%%%%%%%%%%%%%%%%%%%%%%%%%%%%%%%%%%%%%%%%%%%%%%%%

\subsubsection{Near-IR spectroscopy}

The correlation of the line emission variability during the outburst/quiescence stages and the stellar properties of the
source is crucial to understand the physical mechanism(s) active
during the EXor evolutionary phase.

By examining the near-IR EXor spectra appeared sofar in the
literature (see Table~\ref{spectra:tab}), a systematical study of the emission line variability results unaffordable in terms
of neither intensity nor profile. This derives from both the lack of
any spectroscopical monitoring and the variety of different ways
exploited to obtain, reduce and (un)calibrate the spectra.
Nevertheless two exceptions exist: two objects (XZ Tau and PV
Cep) show significant Br$\gamma$ flux variations (by a
factor of two or more) on timescale of years or less. In XZ Tau,
F(Br$\gamma$)= 2.1 $\pm$ 0.6 10$^{-13}$erg s$^{-1}$cm$^{-2}$ (Nov.
1983 - Evans et al. 1987), 0.42 $\pm$ 0.06 and 0.9 $\pm$ 0.1
in the same units, in 1986-87 and 1988-89, respectively  (Carr
1990; Giovanardi et al. 1991). In PV Cep, while F(Br$\gamma$)
varied from 0.44 $\pm$ 0.05 (Jul. 1986 - Carr 1990) and 0.8
$\pm$ 0.1 (Oct. 1986 - Carr 1990), F$_{H_2}$(2.12$\mu$m)
remained constant (0.33 $\pm$ 0.05 and 0.36 $\pm$ 0.1) in the same
two occasions. This result thus suggests that HI recombination
should be more prompt than molecular material to response to
accretion events, or even more adequate to trace them.
Unfortunately, XZ Tau and PV Cep were not observed photometrically
in that periods, so that any correlation between spectral
behaviour and brightness status is prevented.

Near-IR spectra of all the EXors present similar features:
they are emission line spectra usually dominated by the hydrogen
recombination (Brackett and Paschen series), which signals the
presence of ionized gas close to the star. CO overtone emission
v=2-0, v=3-1 is commonly detected along with weaker atomic features
(e.g. MgI at 1.503 $\mu$m and NaI at 2.208 $\mu$m). The
observed spectra are much more similar to those of the accreting T
Tauri stars (Greene \& Lada 1996) than the FUor ones (Reipurth \&
Aspin 1997), apart a couple of exceptions.

Remarkably, some photospheric absorption lines (mainly atomic)
detected in EXor and other young accreting objects, that are used to
estimate the near-IR veiling, are sometimes observed in emission.
Such circumstance, already noted in Paper I, has been interpreted as
a clue that chromospheric emission lines more than a continuum
source are the cause of the veiling (Fohla \& Emerson 1999).

\subsubsection{Near-IR colours vs. K magnitude}

\begin{figure}
 \centering
   \plotone{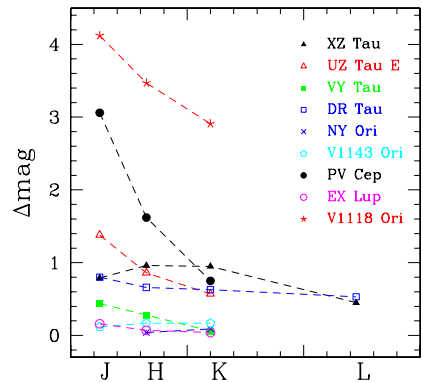}
   \caption{Maximum brightness variations of the individual sources according to the observations
   in Table~\ref{nir:tab}.
   \label{deltamag:fig}}
\end{figure}

The aim of this section is to investigate whether or not a
recognizable trend exists among the EXors to become bluer (or
redder) while they dim; in other words to present a variable SED (in
the near-IR) moving from a flat to a more peaked one (or vice versa)
while the source is progressively fading. The answer should help in
clarifying the mechanism(s) which are active during the outburst
phases. The analysis of individual objects (e.g. Kenyon et al. 1994;
Lorenzetti et al. 2006) allows to recognize a trend toward the
blueing. Now, to provide a statistically significant sample of
data, we use the catalog listed in Table~\ref{nir:tab} to construct
near-IR colour vs. magnitude diagrams. All of them show a
similar behaviour, so we give in Figure~\ref{J-K:fig} the [J-K] vs.
K plot as a typical example. Additionally, Figure~\ref{deltamag:fig}
depicts for each source the magnitude difference ($\Delta$ =
mag(min) - mag(max)), as a function of the wavelength, between the
two epochs corresponding to the maximum and minimum flux in the J
band (arbitrarily chosen). Bias effects could be introduced if
different brightness variations were attributable to substantially
different time-scales for the various sources, but looking at the
dates of the observations (Table~\ref{nir:tab}), the sampling times
seem quite similar (typically months, years), thus the occurrence of
peculiar time-scales can be ruled out. \\

From the data depicted in Figures~\ref{J-K:fig} and
\ref{deltamag:fig} some conclusions can be derived for the EXors:

{\it (i)} from Figure~\ref{deltamag:fig} we see that a substantial
brightness variation ($\Delta$mag above the 1mag level) has been
clearly observed only in two sources, namely V1118 Ori and PV Cep,
and marginally in UZ Tau E. Such result is a direct consequence
of the random plans of the IR observations: indeed, while V1118 Ori
has been monitored to probe brightness variations, all the other
sources were not observed systematically, hence the observed photometric variation ($\Delta$mag $<$ 1) represents just a lower limit to the intrinsic amplitude.

The presented observational scenario, being a direct consequence of
the random plans of the IR observations, does not reflect any
peculiar property of that couple of sources; indeed, all the EXors
underwent large ($\Delta$mag $>$ 4) optical variations (see
Table~\ref{parameters:tab}) typical of the class, that were not
monitored the IR band.

{\it (ii)} Irrespectively of the source brightness (from 12 to 7
mag in K band, see Figure~\ref{J-K:fig}) the amplitude of the
[J-K] colour variations is substantially similar (about 1 mag) for
all the sources with respect to a mean value of about 1.5-2.0 mag:
PV Cep, sometimes also classified as an Herbig or a FUor
star, represents the only exception.

{\it (iii)} EXors do not behave all in the same way: while becoming
fainter (see Figure~\ref{J-K:fig}), some sources tend to become
bluer or not to exhibit a definite colour trend (V1143 Ori, EX Lup,
DR Tau, XZ Tau, VY Tau);  while others become definitely
redder (V1118 Ori, PV Cep, UZ Tau E). Noticeably, these are the
same three sources for which the full amplitude of the IR variation
has been observed (see above). Thus, it seems that only during the
outburst phase a peak at a relatively high temperature becomes more
and more evident (blueing). In the other sources colder
contributions typical of a disk temperature stratification prevail.

{\it (iv)} extinction variations are to be ruled out since the very
low values of A$_V$ (1-2 mag, see Table~\ref{parameters:tab}).

Summarizing, the present view tends to identify the EXOrs
appearance during the active phase with that of the classical
T-Tauri stars. This issue will be discussed in more detail in
Sec.5.1.4. at the light of the EXor complete SED's, including the
mid- and far-IR contributions, as well.

%%%%%%%%%%%%%%%%%%%%%%%%%%%%%%%  FIGURE deltamag vs lambda %%%%%%%%%%%%%%%%%%%%%%%%%%%%%%%%%%%%%%%%%%%%
%%%%%%%%%%%%%%%%%%%%%%%%%%%%%%%%%%%%%%%%%%%%%%%%%%%%%%%%%%%%%%%%%%%%%%%%%%%%%%%%%%%%%%%%%%%%%%%%%%%%%

\subsubsection{N band data}

The observations in the N band (8-13 $\mu$m) given in
Table~\ref{mir:tab} are quite sparse and almost never simultaneous
with the near-IR ones. Nevertheless two facts can be pointed out:

1) in few occasions, K and N band observations can be considered
quasi-simultaneous (in the sense that both are carried out within
one month) with respect to the time scale of the EXor variations. By
comparing $\Delta$K and $\Delta$N variations, they appear quite
correlated within the errors.

2) A very crude evaluation of the [K-N] colour indicates values for
all the sources that range between 4 and 7, typical of thick disks.
Remarkably, any EXor exhibits a [K-N] colour between 0 and 1,
which is typical of normal Main Sequence stars. This finding
confirms a known result (e.g. Strutskie et al. 1990), according to
which the transition phase between the thick disk stage and the end
of the active disk survival could be very fast.

\subsubsection{Far-IR data}

\begin{figure}
\epsscale{.80}
 \centering
   \plotone{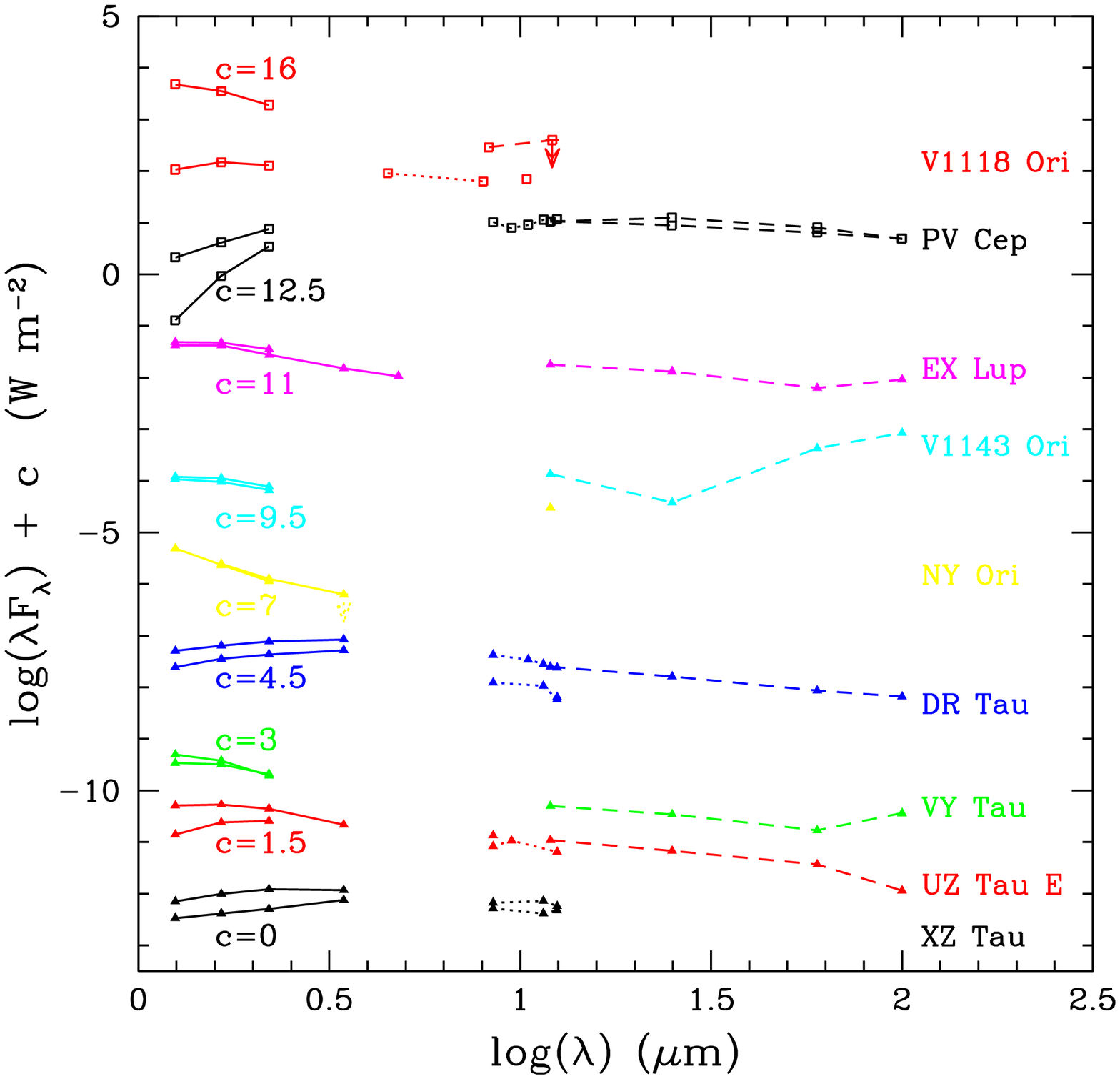}
   \caption{SED's of the EXor sources constructed with the catalog
   data given in Tables~\ref{nir:tab}, ~\ref{mir:tab} and
   ~\ref{fir:tab}. Ground-based observations (solid line in the near- and dotted in the
   mid-IR) refer (when available) to the two different dates when minimum and maximum
   brightness was observed, respectively; dashed line refers to IRAS fluxes.
   When connected, the  data points refer to the same epoch.
   \label{sed:fig}}
\end{figure}

A summary of all the literature far-IR data (12-100 $\mu$m) of EXors
is given in Table~\ref{fir:tab}. Any evaluation for possible far-IR
variability is prevented at the moment, since a single observation
per source is available from the IRAS and MSX database. Only for PV
Cep a second observation (about 15 years later) was obtained by
combining the ISO spectrophotometry with IRAS bandpasses; no
significant variation seems to occur in this wavelength range.
The far-IR data are useful to construct comprehensive SED's over a
wide (2 decades) band where EXors are expected to emit a large part
of their power. These are given in Figure~\ref{sed:fig}: two
determinations, mainly in the near-IR (1-5 $\mu$m solid line),
correspond to the maximum and minimum observed brightness; mid-IR
ground-based observations (8-13 $\mu$m dotted lines) partially
overlap with the IRAS/MSX data (12-100 $\mu$m dashed line); when
connected, the data points refer to the same epoch. The ground-based
mid-IR photometric values, although carried out in different
occasions, are equal to or weaker than the IRAS 12 $\mu$m fluxes,
that could be contaminated by diffuse circumstellar emission. Mid-IR
data are also characterized by a certain degree of variability quite
comparable with the near-IR one. As already commented in Sec.5.1.2,
only the sources V1118 Ori, PV Cep and UZ Tau E present a SED which
tends to peak at shorter wavelength during the outburst, when the
brightness increases. The remaining sources that show some
variations do not change their SED slope. In essence, only during
the active burst phase the EXor SED seems to look like the T Tauri
averaged SED (D'Alessio et al. 1999), while for the remaining time
EXors show a flatter spectral distribution more typical of a disk
temperature stratification.

%%%%%%%%%%%%%%%%%%%%%%%%%%%%%%%  FIGURE sed %%%%%%%%%%%%%%%%%%%%%%%%%%%%%%%%%%%%%%%%%%%%
%%%%%%%%%%%%%%%%%%%%%%%%%%%%%%%%%%%%%%%%%%%%%%%%%%%%%%%%%%%%%%%%%%%%%%%%%%%%%%%%%%%%%%%%%%%%%%%%%%%%%

\subsubsection{Catalog data viewed in a phenomenological framework}

The results derived from the catalog data (see Sect. 5.1.2, 5.1.3
and 5.1.4), although not obeying to any observational
strategy, can be still referred to a comprehensive interpretation.
Indeed, an active disk can show two types of instabilities:
dynamical and thermal. While the former (Adams \& Lin 1993) are
unlikely responsible for the observed time-scale of the photometric
variations, the latter (Hartmann, \& Kenyon, 1985) offer a more
acceptable explanation. The disk accretion hypothesis is based on
the viscous friction between particles with different velocities.
The efficiency of this coupling is an increasing function of the
temperature: in fact a local increase of the temperature
augments the viscosity, the efficiency of transferring angular
momentum, the accretion rate and, again, the disk temperature.
Depending on the characteristics of the initial perturbation, it
will propagate along the whole disk or will become exhausted before.
According to this scheme, the predicted IR spectral shape of the
continuum emitted by the disk is characterized by an initial
brightening at a given wavelength which is different depending on
the distance (from the disk center) at which the perturbation
onsets. By assuming an onset around 10 $\mu$m, the source tends to
become initially redder, but while the propagation is going toward
the center, the source becomes bluer and bluer and the power emitted
at smaller wavelengths becomes progressively larger. This picture
agrees quite reasonably with the observational evidences: only
when the outburst phase is fully sampled we are able to see the
blueing while brightening, otherwise we can see just smaller
amplitude variations showing a barely defined (bluer or redder)
trend.

\subsection{New results on V1118 Ori}

\subsubsection{Near-IR photometry}

The near-IR photometric data listed in Table~\ref{mag:tab} and shown
in Figure~\ref{lightcurve:fig} agree quite well with the light curve
of the outburst recently monitored in the V band (Garcia, Parsamian
\& Velazquez 2006). Our data indicate that the post-outburst phase
has been sampled only during the short period between October and
November 2005 (Figure~\ref{lightcurve:fig}), characterized by
irregular fluctuations without any superposition of an appreciable
fading trend. The further declining down to the lowest level of IR
intensity has remained undetected, while the subsequent monitoring
between September and December 2006 (bottom part of
Table~\ref{mag:tab} and right panels of Figure~\ref{lightcurve:fig})
has been performed once the quiescent phase was already reached
(likely around May-June 2006, as deduced by extrapolating the visual
light curve). The near-IR monitoring allows us to state some
observational facts: {\it (i)} the quiescent phase of V1118 Ori
corresponds to mag levels of about 12.7, 11.7 and 11.0, in the bands
J, H and K, respectively. {\it (ii)} The dispersion around the
quiescence mean value is significantly smaller (e.g. 0.04 mag in the
J band) with respect to the value (0.20 mag) derived during the
previous fading phase. This circumstance suggests that
quiescence is a much more stable phase than declining, which
is accompanied by significant and irregular magnitude variations
(0.5 mag or more, see left panels of Figure~\ref{lightcurve:fig}) on
a daily time-scale. Such a behaviour signals a reduced level of
activity during the quiescence and hence the existence of some
correlation between the mechanism(s) responsible for both the
outburst and the short time-scale activity. If, as widely accepted,
the accretion rate variation regulates the outburst occurrence, then
our results support the hypothesis that this mechanism also
regulates the irregular and rapid variations and it is indeed
dynamical on (at least) two time-scales, namely months and days.
{\it (iii)} The colours [J-H] and [H-K] corresponding to the
quiescent phase have reached once again their maximum values of 1.0
and 0.7 mag, respectively, already indicated as typical of the
quiescence during the year 2000 (Paper I). This means that V1118 Ori
has a periodical behaviour, namely tends to show always the same
colours in the same phase, even if in different epochs. Moreover,
these observations definitely confirm that the object becomes redder
when fading (as already indicated in Figure~\ref{J-K:fig} and
discussed in Sect.5.1.2).

\subsubsection{Near-IR spectroscopy}

The real asset among this data set collected after Paper I is
certainly represented by the third near-IR spectrum, taken in 2006
September, namely one year later than the first two. As illustrated
in Figure~\ref{nirspectra:fig} (green plot) it does not show (
at our sensitivity) the presence of any emission line, as if V1118
Ori were apparently become a featureless object once its quiescent
phase has been reached. Such a behaviour is confirmed by the fourth
spectrum taken 18 days later (blue plot). By considering the maximum
line flux detected in the first spectrum within the range 1.44 -
2.35 $\mu$m, namely the value 1.2 $\pm$ 0.2 10$^{-13}$erg
s$^{-1}$cm$^{-2}$ relative to the Br$\gamma$ transition (see Paper
I), and the more recently detected 3$\sigma$ upper limit (0.2
10$^{-13}$erg s$^{-1}$cm$^{-2}$) measured at the Br$\gamma$
frequency, we obtain that the line emission has faded by more
than a factor of 6. The only spectral feature still
present in the third spectrum is the 2.3 $\mu$m CO band (2-0) now
seen in absorption. This feature is totally absent in the fourth
spectrum. This occurrence is surprising since the first two spectra
taken one year before, when the source was brighter of about 1.5
mag, presented the same features, but in emission at a S/N level of
4 (see Figure~\ref{nirspectra:fig}).\\

A couple of similar cases of spectral variations occurred in more
embedded YSO's, not related to external phenomena (e.g. variable
extinction or binary occultation), but indicative of changes in the
physical conditions of the star and its circumstellar environment,
have been discussed in the literature. The source NGC2024 IRS2 had a
burst of its IR continuum of about one magnitude and a simultaneous
brightening in HI recombination line intensities together with a
large velocity shift (50-70 km~s$^{-1}$) in the line emission
(Nisini et al. 1994). More recently, different outburst phases of
the source V1647 Ori have been monitored by means of near-IR
spectroscopy (Gibb et al. 2006), demonstrating how the post-outburst
phase is characterized by a declining temperature of the hot CO gas
formed in the inner part of the disk, and by a substantial decrease
of the fast
wind.\\

The disappearing of all the emission lines at our
sensitivity is indeed quite dramatic and remains a very rare case
of a directly documented spectral variation. Nevertheless, such an
event can be easily accounted for in the currently accepted
framework which describes the EXors phenomenology.

The recurrent outbursts are interpreted (see also Sect.5.1.5) as
caused by variable mass accretion. A suddenly increase of the
accretion rate forces the disk to re-radiate with a consequent
brightening of the IR continuum. The same accretion event
regulates also the mass loss rate which is strictly related to the
accretion rate (e.g. Shu al. 1994); it determines an intermittent
mass loss monitored by observing the alternate appearance and
disappearance of emission lines and molecular (CO) bands that
originate, respectively, in the ionized and neutral component of the
stellar wind. Concerning the CO variability, Biscaya et al. (1997)
investigated this topic and their results indicate that the majority
of the YSO's with CO bands in emission are variable on time-scales
as short as a few days (DG Tau even periodically). Therefore, in
this respect, V1118 Ori does not represent an exception, although
our data tend to exclude CO variability on short time-scales. A
systematical monitoring of the CO variability has been never
afforded to date, however, Biscaya et al. (1997) consider the
different models able to account for CO emission vs. absorption. All
of them are based on accretion processes on circumstellar disk, but
their predictions are often conflicting: the spectra presented here
tend to support those models where an increasing accretion rate may
lead to a significant increase of UV radiation in the accretion
shock that will heat the surface layer
of the disk, so favoring CO emission.\\

The shape of the continuum is also remarkable
(Figure~\ref{nirspectra:fig}) since confirms once again (see Sect.
5.1.2) an intrinsic reddening of V1118 Ori as it fades.\\

Finally, the two near-IR spectra of V1118 Ori (taken in 2005) do not
show any clue of line emission variability on 50 days time-scale.
The repeatability of the second near-IR spectrum provided us the
opportunity to identify emission lines that, although present in the
first spectrum as well, were not mentioned in Paper I because
appeared weak and/or blended features. The list of these newly
identified lines is given in Table~\ref{lines:tab} where
the line fluxes (computed on the first spectrum) are given only for
those having S/N \gapprox 3. Similarly to those identified in Paper
I, also these additional features are all characteristic of the
ionic emission component typical of T Tauri stars.

\subsubsection{Polarization}

\begin{figure}
 \centering
   \plotone{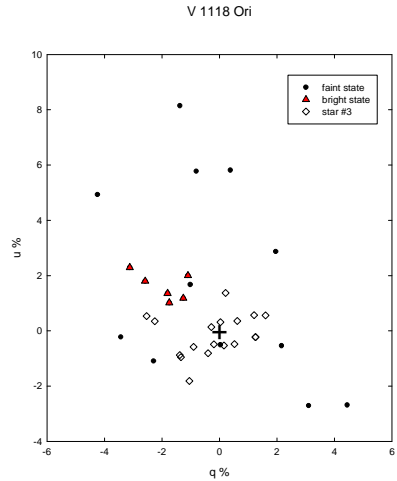}
   \caption{Behaviour of the u vs. q normalized Stokes parameters.
   Different symbols refer to a fainter state of V1118 Ori, a brighter one and a reference star (\# 3).
   Errors on u, q parameters are about 0.5\% for the bright state and about 1\% for the
   faint one. The origin is indicated by a cross.
   \label{polar:fig}}
\end{figure}

The results of our polarimetric monitoring performed in the I band
are given in Table~\ref{polar:tab}; data were also obtained for a
source located 80 arcsec to the North of V1118 Ori (indicated in
that Table as reference star). The behaviour of the normalized
Stokes parameters $(u,q)$ is depicted in Figure~\ref{polar:fig},
where different symbols are related to a fainter state (14-15 mag in
I band), a brighter (12-13 mag) state of V1118 Ori, and the
reference star, which has approximately the same brightness as V1118
Ori in the fainter state.

The difference between V1118 Ori and the reference star is
noticeable. The parameters distribution of the latter is scattered
around the origin, substantially showing a lack of any intrinsic
polarization. On the contrary, the u,q values corresponding to the
brighter state of V1118 Ori are definitely less scattered, certainly
displaced from (0,0) position and give indication of a
substantial polarization (p~=~2.5\% $\pm$ 0.4\% and $\theta$
=~64$\degr$ $\pm$ 4$\degr$). Such polarization is likely intrinsic,
since an interstellar origin is not compatible with the low
reddening (A$_V$ $\sim$ 1-2 mag, see Paper I). Also scattering by
dust lost by the star during previous outbursts tend to be ruled out
by the same argument. According to the previous discussion
(Sect.5.1.5), when the star+disk system fades, it is expected to
come back to the original and stationary pre-outburst status provoking also a wind decreasing and, consequently, we can see the
real heavily spotted and magnetized photosphere. This could be the
reason for having detected high and variable values of polarization
during the fainter state (Figure~\ref{polar:fig}).

Our data suggest that short time-scale variability might be
associated to variations in polarization. Wood et al. (1996)
investigated the photo-polarimetric variability of a magnetic
accretion disk model for pre-main-sequence T Tauri stars. In their
model the matter from the disk accretes along the magnetic field
lines onto the stellar surface producing hot and polarized
spots; stellar rotation causes the photometric and
polarimetric variations; these latter are also caused by the
scattering from the disk. At the present status, a direct comparison
with model predictions is difficult since the behaviour of V1118 Ori
is not so straightforward: rotation might not be the main
responsible for the observed photometric variations; further causes
of scattering (namely of polarization) may exist, as asymmetric
envelopes or accretion disk, and we cannot exclude that remnants of
them are still present during fainter stages, complicating the whole
picture.
\clearpage
\onecolumn

%%%%   TABLE - LINES  %%%%%%%%%%%%%%%%%%%%%%%%%%%%%%%%%%%%%%%%%%%%%%%%
\begin{deluxetable}{ccc}
\tabletypesize{\footnotesize}
\tablecaption{ Additional (see Paper I) line emission identification
of V1118 Ori\label{lines:tab}}
\tablewidth{0pt}
\tablehead{
\colhead{$\lambda_{vac}$} & \colhead{Ident.}& \colhead{F $\pm$ $\Delta$F}
}
\startdata
($\mu$m)       &     & (10$^{-13}$erg s$^{-1}$cm$^{-2}$)\\
\tableline
1.1290   &  OI      &  0.8 $\pm$ 0.2  \\
1.1756   &  CI      &  0.6 $\pm$ 0.2  \\
1.1831   &  MgI     &  0.5 $\pm$ 0.2  \\
1.1972   &  HeI     &  1.0 $\pm$ 0.2  \\
\enddata
\end{deluxetable}
%%%%%%%%%%%%%%%%%%%%%%%%%%%%%%%%%%%%%%%%%%%%%%%%%%%%%%%%%%%%%%%%%%%%%%%%

%%%%   TABLE - POLARIZATION  %%%%%%%%%%%%%%%%%%%%%%%%%%%%%%%%%%%%%%%%%%%%%%%%
\begin{table}
\begin{center}
\tabletypesize{\footnotesize}
\caption{ Polarization data of both V1118 Ori and a reference star
in the I band \newline \label{polar:tab}}
\begin{tabular}{c|ccccc|cccc}
\tableline\tableline
MJD &  \multicolumn{5}{c}{V1118 Ori}  &
\multicolumn{4}{c}{reference star}\\
\tableline
 & I (mag)& p$_I$ (\%)& $\sigma_p$ & $\theta$($\degr$)& $\sigma_{\theta}$& p$_I$ (\%)& $\sigma_p$& $\theta$($\degr$) &
 $\sigma_{\theta}$ \\
\tableline
53689.04  & 11.96  &  2.0 & 0.4  & 75    &  5   &  0.7   &  0.6   & 60   &   20 \\
53689.05  & 11.95  &  3.9 & 0.4  & 72    &  3   &  --    &   --   & --   &   -- \\
53698     & 12.36  &  2.3 & 0.3  & 59    &  4   &  1.1   &  0.3   & 104  &   8  \\
53702     & 12.41  &  1.7 & 0.1  & 68    &  2   &  1.6   &  0.3   & 105  &   6  \\
53703     & 12.49  &  3.1 & 0.3  & 73    &  2   &  0.7   &  0.3   & 160  &   12 \\
53713     & 12.99  &  2.2 & 0.4  & 72    &  4   &  1.5   &  0.4   & 43   &   8  \\
53779     & 13.25  &  3.4 & 0.2  & 92    &  2   &   --   &   --   &  --  &   -- \\
53781     & 13.12  &  8.3 & 0.5  & 50    &  2   &   1    &  1     & 180  &   23 \\
53782     & 12.98  &  1.9 & 0.4  & 60    &  5   &  0.5   &  0.2   & 120  &   11 \\
53785     & 13.39  &  6.5 & 0.3  & 65    &  1   &  2.6   &  0.9   &  80  &   10 \\
53786     & 13.50  &  2.5 & 0.4  & 103   &  5   &   --   &   --   &  --  &   -- \\
53787     & 13.52  &  2.2 & 0.5  & 173   &  6   &  1.3   &  0.3   & 14   &   7  \\
53788     & 13.51  &  3.5 & 0.8  & 28    &  7   &  1.6   &  0.5   & 110  &   10 \\
53789     & 13.31  &  0.5 & 0.4  & 140   &  20  &  0.9   &  0.3   & 120  &   10 \\
54029     & 14.69  &  6   & 1    &  43   &  7   &  1.7   &  0.3   & 9    &   6  \\
54038     & 14.55  &  6   & 1    & 49    &  7   &  0.3   &  0.4   & 40   &   30 \\
54085     & 14.59  &  5.2 & 0.6  & 164   &  3   &  0.7   &  0.1   & 195  &   6  \\
54086     & 14.61  &  4.1 & 0.8  & 159   &  5   &  2.5   &  0.3   &  72  &   3  \\
\tableline
\end{tabular}
\end{center}
\end{table}
%%%%%%%%%%%%%%%%%%%%%%%%%%%%%%%%%%%%%%%%%%%%%%%%%%%%%%%%%%%%%%%%%%%%%%%%

\newpage
\twocolumn

%%%%%%%%%%%%%%%%%%%%%%%%%%%%%%%  FIGURE POLARIZATION %%%%%%%%%%%%%%%%%%%%%%%%%%%%%%%%%%%%%%%%%%%%
%%%%%%%%%%%%%%%%%%%%%%%%%%%%%%%%%%%%%%%%%%%%%%%%%%%%%%%%%%%%%%%%%%%%%%%%%%%%%%%%%%%%%%%%%%%%%%%%%%%%%

\section{Concluding remarks}

We have compiled the first catalog of all the IR (1-100 $\mu$m)
photometric and spectroscopic observations of EXors appeared in the
last 30 years literature. Analyzing the catalogued data the
following can be summarized:

\begin{itemize}
\item The constructed database shows that the 9 EXors have been photometrically observed in the 1-10
$\mu$m range on only 88 occasions, apart a daily monitoring of DR
Tau. A (quasi-)simultaneous entire SED in that range has never been
observed, but in 6 occasions a partial coverage (typically KLMN
bands) has been obtained. Reference is given to the far-IR
photometry obtained with IRAS/MSX/ISO and all the available mid- and
far-IR spectroscopy is also mentioned. Although the IR monitoring of
EXors conducted so far does not obey to any observational strategy,
some conclusions on the general behaviour of these sources can be
drawn.
\item Spectroscopically, two objects present significant Br$\gamma$ flux
variations on timescale of years or less and all EXors tend to show
line emission spectra dominated by hydrogen recombination, where CO
overtone emission is commonly detected with weaker atomic features.
EXor IR spectra resemble those of the accreting T Tauri stars more
than those of FUor objects.
\item The objects showing larger ($\Delta$mag $>$ 1) flux variations,
namely those where we are reasonably confident to have monitored the
full amplitude of the IR variation, become definitely bluer when
brightening, and such property can be assigned to the whole class.
Extinction variations can be ruled out because of the low A$_V$
values (1-2 mag).
\item Near- and mid-IR variations appear correlated and the [K-N]
colour confirms that, for the EXors as well, the transition phase
between the thick disk stage and the end of disk survival could be
very fast.
\item Far-IR (12-100 $\mu$m) variability cannot be evaluated since a single observation per source
exists at the moment; one source was observed twice on a timescale
of 15 yrs and does not show any significant variation. However,
far-IR data are essential to construct comprehensive SED's over a
wide (2 decades) band. These SED's tend to peak at shorter
wavelengths during the active burst phase, namely while brightening
they look like the T Tauri averaged SED; the remaining time EXors
show a flatter SED more typical of a disk temperature
stratification.
\item The observational evidences can be coherently interpreted in
the framework of the disk accretion hypothesis based on a viscous
friction between particles with different velocities. A local
increase of the accretion along the disk causes thermal
instability. Depending on the characteristics of the initial
perturbation it will propagate along the whole disk or will become
exhausted before. Therefore the initial brightening can occur at
any wavelength, causing a bluer o redder SED, but while the
propagation is going toward the center the source appear bluer and
bluer.
\end{itemize}

Moreover, our monitoring program on the EXor variable V1118 Ori is
still going on, and the following results complement those presented
in Paper I:

\begin{itemize}
\item Near-IR (JHK) photometric monitoring has covered part of the
post outburst phase and the quiescent phase. The former is
characterized by irregular fluctuations, whose amplitude is
significantly  reduced (from 0.5 mag to 0.04 mag) during the latter
phase. This testifies the existence of some correlation between the
mechanism(s) responsible for both the outburst and the short
timescale variability. Being accepted that the accretion rate
variation regulates the outburst, our observations indicate that
such accretion rate is dynamical on (at least) two timescales
(months and days).
\item A second near-IR spectrum (0.83-2.35 $\mu$m) taken 50 days
after the first one (Paper I) does not show any clue of line
emission variability, but provides us a cross-confirmation of
features not mentioned before because appeared weak or blended.
These additional features are all characteristic of the ionic
emission component typical of T Tauri stars.
\item A third near-IR spectrum, taken one year later, during the
quiescent phase, surprisingly lacks of any spectral feature at
our sensitivity except the CO (2-0) band that appears 
now as a weak absorption instead of emission. This observational
circumstance, although very rarely detected, agrees quite naturally
with the interpretation of the EXor events as due to variable
accretion.
\item The lack of molecular component of shock excitation (H$_2$
emission) is also given by the analysis of the archival {\it
Spitzer} images in the 4.5 $\mu$m channel. {\it Spitzer} data,
complemented with ground-based N-band photometry (10.4 $\mu$m),
constitute the first mid-IR detection of V1118 Ori.
\item Finally, the first polarimetric observations (in the I band)
indicate that V1118 Ori is intrinsically polarized (at a level
of 2.5\%). Moreover, higher and more variable values of polarization
are observed during the fainter state suggesting that when the
envelope is expected to partially disappear, we can see the real
heavily spotted and magnetized photosphere.
\end{itemize}

\end{document}